\input epsf.sty
\documentstyle[epsf,editedvolume]{crckapb} 
\begin{opening}
\title{THE QSO EVOLUTION DERIVED FROM THE {\it HBQS} \protect\\
       AND OTHER COMPLETE QSO SURVEYS}
\author{F. LA FRANCA}
\institute{Dipartimento di Fisica, Universit\`a degli studi ``Roma TRE''\\
           Via della Vasca Navale 84, I-00146, Roma, Italy}
\author{S. CRISTIANI}
\institute{Dipartimento di Astronomia, Universit\`a degli studi di Padova \\
Vicolo dell'Osservatorio 5, I-35122 Padova, Italy}

\end{opening}
\runningtitle{The QSO evolution from the {\it HBQS}}

\begin{document}

% The \begin{document} command comes after the \end{opening}
% command.

\section{Introduction}

In this paper we present a new determination of the QSO Luminosity Function
(LF) and its evolution, based on new data coming from the Homogeneous Bright
QSO Survey ({\it HBQS}, partly published in Cristiani {\it et al.} 1995). We
also take into account the most relevant already existing surveys. 

The {\it HBQS}
was started in 1989 in the framework of an ESO Key-programme. The
survey covers a total of 555 deg$^2$ subdivided in 22 ESO or UKST fields at
high galactic latitude around the south galactic pole. 
Two Schmidt plates for each bandpass $U$, $B'$ or $B_J$, $V'$, $R$ or $OR$ and
$I$ have been obtained at the UKST or ESO Schmidt telescope, usually within a
few months interval, in order to minimize the effects of variability. The plate
material has been scanned on the COSMOS microdensitometer. 
According to the ESO/UKSTU numeration, the 22
fields which have been observed are: 287, 290, 295, 296, 297, 349, 351, 355,
406, 407, 408, 410, 411, 413, 468, 469, 470, 474, 479, 534, SA94, SGP. 

We have selected as candidates all the UVx ``not-extremely-extended'' objects
satisfying a type of {\it modified Braccesi less-restricted} two-color
criterion (La Franca, Cristiani \& Barbieri 1992). The magnitude intervals in
which the selection is virtually complete vary from field to field between
$15.0<B<17.3$ and $15.0<B<18.8$. The sample extracted from these data includes
327 QSOs ($M_B<-23$, $q_0 = 0.5$, and $H_0=50~Km/s/Mpc$). 

The other samples we have used for the computation of the QSO LF are:
A) The Edinburgh QSO Survey (EQS, Goldschmidt
{\it et al.} 1992), which covers $333$ deg$^2$ in the interval $15.0<B<16.5$; B)
The Large Bright QSO Survey (LBQS, Hewett, Foltz and Chaffee 1995), which covers
$454$ deg$^2$ in the interval $16.5<B<18.5$; C) The SA94 Survey (La Franca {\it
et al.} 1992), which covers $10$ deg$^2$ in the interval $15.0<B<19.9$; D) The
Durham/AAT Survey (Boyle {\it et al.} 1990), which covers $11.9$ deg$^2$ in the
interval $16.0<b<20.9$; E) The ESO/AAT (Boyle, Jones and Shanks 1991), which
covers $0.85$ deg$^2$ from $J=18.0$ down to $J=21.75-22.0$; F) The (ZM)$^2$B
Survey (Zitelli {\it et al.} 1992), which covers an area of 0.69 deg$^2$ down to
$J=20.85$ and 0.35 deg$^2$ down to $J=22.0$. 

\begin{figure}
%\vspace{6cm}
\epsfysize=60truemm
\epsffile{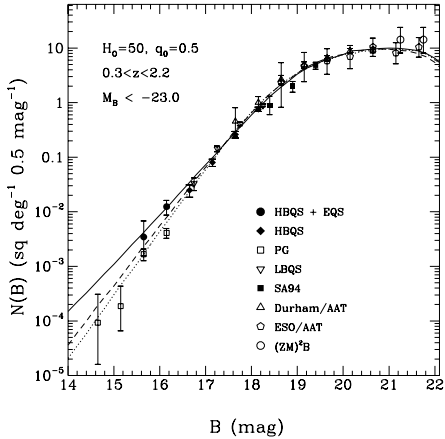}
\caption{The QSO number counts. Continuous line: LDLE model C. Dashed line:
PLE model B. Dotted line: PLE model A (see Table 1).}
\end{figure}

\section{The QSO Counts}

The B number-magnitude counts, in the interval $0.3<z<2.2$, for all the samples
used in the computation of the optical LF are shown in Fig. 1. As first
indicated by Goldschmidt {\it et al.} (1992), for magnitudes brighter than
$B=16.4$ the QSO surface density turns out to be a factor 2.5 higher than what
measured by the PG survey (Schmidt \& Green 1983). The data from the {\it HBQS}
collect 7 QSOs at magnitudes brighter than $B =16.4$, corresponding to a
surface density of $0.013^{+.007}_{-.005}~deg^{-2}$. With the addition of the
EQS the surface density at $B<16.4$ becomes $0.016 \pm 0.005 ~ deg^{-2}$,
corresponding to a total of 14 QSOs over an area of $888~ deg^{2}$. In a larger
area ($\sim 6800$ deg$^2$), the PG survey collects 41 QSOs corresponding to a
QSO surface density of $0.006 \pm 0.001~deg^{-1}$. 

\section{The Luminosity Function}

In the interval $0.3<z<2.2$ the QSO LF has
been usually parameterized with a double power law Pure Luminosity Evolution
(PLE) model\hfill\break 
\vskip 0.5mm
\centerline{$ \Phi (M_B,z) = { {\Phi^\ast} \over {
10^{0.4[M_B-M_B(z)](\alpha + 1)} + 10^{0.4[M_B-M_B(z)](\beta + 1)} } }, $}
\vskip 2mm
\noindent
where $\alpha$ and $\beta$ correspond to the faint-end and bright-end slopes of
the optical LF, respectively. With this parameterization the evolution of the
LF is uniquely specified by the redshift dependence of the  break magnitude, $
M_B (z) = M_B^{\ast} - 2.5k\log(1+z)$ corresponding to a power law evolution in
the optical luminosity, $L^{\ast}\propto (1+z)^k$ (see Boyle, Shanks,
Peterson 1988). 

In order to compute the QSO LF, we have simulated, via Monte Carlo techniques,
the effects of the various stochastic processes that in reality contribute
to determine the observed magnitudes and colors.

The apparent B magnitudes of the QSOs have been computed following the
K-correction according to Cristiani \& Vio (1990), and computing the average
galactic absorption $A_B$ according to the HI maps:\hfill\break \centerline{$ B
= M_B + 5 \log [A(z)c/H_0] + 25 + K(z) + A_B + \epsilon(B)$} where
$\epsilon(B)$ simulates a noise with Gaussian distribution with variance $
\sigma^2_B = \sigma^2_{phot}(B) + \sigma^2_{B,\gamma}.$ $ \sigma^2_B$ takes
into account both the photometric errors of the survey $\sigma_{phot}(B)$ and
the apparent magnitude dispersion due to the QSO spectral slope dispersion
$\sigma_{B,\gamma} = 2.5\sigma_{\gamma} \log (1+z).$

The apparent colors have been generated following the average QSO
colors-redshift dependence $F_{1,2}(z)$ (Cristiani \& Vio 1990). Between each
couple of bandpasses centered at $\lambda_1$ and $\lambda_2$ the resulting
apparent color is: $ C_{1,2} = F_{1,2}(z) + \epsilon_{F_{1,2}}$ where
$\epsilon_{F_{1,2}}$ simulates a noise with Gaussian distribution with variance
$ \sigma^2_{F_{1,2}} = \sigma^2_1 + \sigma^2_2 + \sigma^2_{F,\gamma}.$ $
\sigma^2_{F_{1,2}}$ takes into account the photometric errors of the two
bandpass $\sigma_1$ and $\sigma_2$ respectively, and the QSO color dispersion
due to the spread $\sigma_{\gamma}$ in the slope of the spectra
$\sigma_{F,\gamma} = 2.5\sigma_{\gamma} \log {\lambda_2 \over \lambda_1}.$ The
flux limits and selection criteria of the survey have been applied to the
resulting magnitudes and colors in order to select the ``observed'' QSOs. 

\section{Results and Discussion}

A total sample of 1022 QSOs has been used. The fits have been carried out by
minimizing the $\chi^2$ statistics derived from the comparison of the observed
$(B, z)$ distribution with 2000 simulations of each theoretical LF model (
see Table 1). The significance of
the fitting has also been tested for goodness-of-fit in the interval
$0.3<z<0.6$, in which the fitting probabilities have been computed using the
2D-KS test. We first focus our discussion on a $q_o =0.5$
universe, we will later see how our results change for $q_o = 0.1$. 

\begin{figure}
\epsfysize=60truemm
%\vspace{6cm}
\epsffile{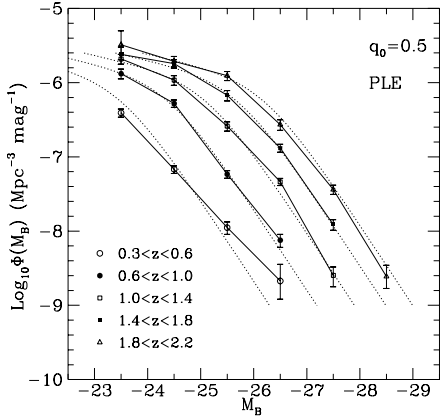}
\caption{The QSO LF. Dotted lines: best fit PLE, model B (see Table 1).}
\end{figure}

In Fig. 2 our best fit PLE model (B) is shown. 
Our PLE model B gives a satisfactory fit of the ($B, z$) distribution with a
global $\chi^2$ probability of 0.21. The PLE model A by Boyle (1992), compared
with the present data, provides a lower but still acceptable probability of
0.12, however with one additional parameter, $z_{cut}=1.9$, representing the
redshift at which the luminosity evolution ``switches off''. At redshift
greater than $z_{cut}$ no further luminosity evolution takes place. Our B model
has a flatter bright slope $\beta$ ($-3.7$ compared to $-3.9$) than the A
model. The evolution parameter $k$ is also smaller ($3.26$ compared to $3.45$).

This difference in the parameter $\beta$ is originated by a larger density of
luminous QSOs (expecially at $z<0.6$) in comparison with the previous data
derived from the PG sample (compare Fig. 2 with Fig. 3b of Boyle 1992). The
higher evolutionary rate $k$ determined by Boyle (1992) is a result of the
introduction of the redshift cut off at $z_{cut}=1.9$, which with our data
results unnecessary.
\begin{table*}
 \centering
 \begin{minipage}{140mm}
  \caption{``Best fit'' parameters for luminosity function models.}
\begin{tabular}{c c c c c c c c c}
&\\
\hline
Model & $\Phi$     & $k_1$ & $k_2$ & $\alpha$ & $\beta$ & $M^{\ast}$ & $\chi^2$ & 2DKS \\
      & $mag^{-1} Mpc^{-3}$ &       &       &          &         &    &$z<2.2$&$z<0.6$\\
\hline
A & $6.5\times 10^{-7}$  & $~~3.45$ &...  & $-1.5~$ & $-3.9~$ & $-22.4$ & $0.12~$ & $0.01$\\
B & $1.1\times 10^{-6}$  & $~~3.26$ &...  & $-1.39$ & $-3.72$ & $-22.3$ & $0.21~$ & $0.02$\\
C & $9.8\times 10^{-7}$  & $~~3.33$ &$~1.2$& $-1.45$ & $-3.76$ & $-22.3$ & $0.53~$ & $0.09$\\
D & $1.0\times 10^{-6}$  & $~~3.29$ &...  & $-1.39$ & $-3.69$ & $-22.4$ & $0.14~$ & $0.03$\\
E & $1.1\times 10^{-6}$  & $~~3.29$ &...  & $-1.39$ & $-3.72$ & $-22.3$ & $0.18~$ & $0.02$\\
F & $1.1\times 10^{-6}$  & $~~3.27$ &...  & $-1.38$ & $-3.72$ & $-22.3$ & $0.18~$ & $0.01$\\
G & $8.5\times 10^{-7}$  & $~~3.19$ &...  & $-1.53$ & $-3.87$ & $-22.4$ & $0.13~$ & $0.01$\\
H & $1.2\times 10^{-6}$  & $~~3.00$ &...  & $-1.37$ & $-3.81$ & $-22.4$ & $0.05~$ & $0.01$\\
I & $4.5\times 10^{-7}$  & $~~3.55$ &...  & $-1.6~$ & $-3.8~$ & $-22.6$ & $0.004$ & $0.01$\\
L & $3.4\times 10^{-7}$  & $~~3.44$ &...  & $-1.57$ & $-3.75$ & $-22.9$ & $0.36~$ & $0.08$\\
M & $4.0\times 10^{-7}$  & $~~3.45$ &$~0.9$& $-1.50$ & $-3.74$ & $-22.8$ & $0.49~$ & $0.11$\\
\hline
& 1$\sigma$ errors:  & $\pm0.07$&$\pm0.3$&$\pm0.07$&$\pm0.13$&$~\pm0.2$&  & \\
\hline
\end{tabular}

NOTES: For $q_0=0.5$: A) Boyle 1992; B) PLE; C) LDLE; 
D) PLE without {\it HBQS}; E) 

PLE without EQS; F) PLE without LBQS; 
G) PLE with $\sigma_\gamma = 0.3$ H) PLE with

$\sigma_\gamma = 0.5$. 
For $q_0=0.1$: I) Boyle 1992; L) PLE; M) LDLE. 

\end{minipage}
\end{table*}

Both A and B models provide an inadequate simulation of the distributions of
the observed data for the low redshift domain, $0.3<z<0.6$. For this subsample
the 2D KS statistics test rejects model A at the 0.01 level and model B at 0.02
level. At magnitudes brighter than $M_B=-25$, in the interval $0.3<z<0.6$, 32
QSOs are observed, while 16 and 19 QSOs are predicted by model A (a $4~\sigma$
discrepancy) and B (a $3~\sigma$ discrepancy) respectively. But as the low
redshift ($z<0.6$) subsample contain only 5 per cent of the complete data set,
in a global comparison of the whole data sample in the interval $0.3<z<2.2$,
the models follow the evolution of the larger fraction of QSOs at higher
redshift, allowing the $\chi^2$ probability of the global fit to reach
satisfactory levels.
No significant difference is obtained by fitting the data excluding each of the
three bright samples ({\it HBQS}, EQS and LBQS) in turn (models D, E and F).

It is interesting to see how the inclusion of a spread in the theoretical
average QSO spectral slope modifies the best fit luminosity function. As
expected (Giallongo and Vagnetti 1992) larger values of the spread of the slope
correspond to lower luminosity evolution parameters and steeper luminosity
functions (model G with $\sigma_\gamma = 0.3$, and model H with $\sigma_\gamma
= 0.5$).

The best description of the observed data has been obtained by decreasing at
low redshift the luminosity evolution of the bright QSOs. This has been
obtained by including a dependence on the redshift and absolute magnitude of
the evolution parameter $k$ such as:\hfill\break
${ }~~~~~~for~M_B \leq M^{\ast}:~ k = k_1 + k_2 (M_B-M^{\ast})e^{-z/{.40}}$\hfill\break
${ }~~~~~~for~M_B > M^{\ast}:~ k = k_1$\hfill\break\noindent
where $M^{\ast}$ is the magnitude of the break in the two power law shape of
the LF. This luminosity dependent luminosity evolution (LDLE) model (model C in
Table 1) has resulted in a better fit of the data (see Fig. 3) giving a
$\chi^2$ test probability of 0.53 in the whole $B, z$ plane, and an acceptable
2D KS probability of 0.09 in the redshift interval $0.3<z<0.6$. With this
model, in this redshift  interval and for magnitudes brighter than $M_B=-25$,
29 QSOs are expected in comparison with the 32 observed.

\begin{figure}
%\vspace{6cm}
\epsfysize=60truemm
\epsffile{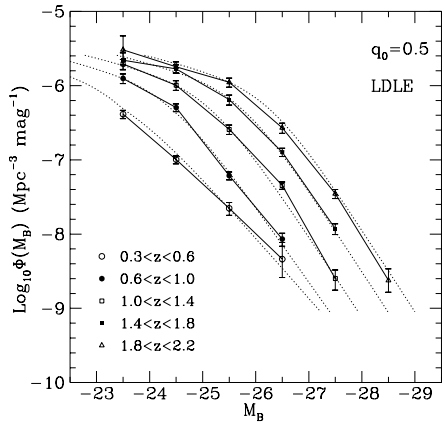}
\caption{The QSO LF. Dotted lines: best fit LDLE, model C (see Table 1).}
\end{figure}

In this way, as shown in Fig. 1, the LDLE reproduces the higher counts of bright
QSOs at $B<17.0$, discovered by the {\it HBQS} and EQS surveys,
much better than the PLE models.
Adopting a $q_o = 0.1$ Universe, as shown in Table 1, the fitting the data with
the PLE model (I) by Boyle (1992) is rejected at 0.004 confidence level. Our
PLE model (L) obtains an acceptable global representation of the data ($\chi^2$
probability 0.36), slightly better than the PLE fit in a $q_o = 0.5$ Universe.
However, in the $0.3<z<0.6$ redshift range the PLE still underestimates the
expected number of QSOs, although the discrepancy is less serious than in the
$q_o = 0.5$ case: 44 QSOs are observed in this domain versus 33 predicted in
our L model (24 in the I model). The introduction of a LDLE parameterization
provides satisfactory representation of the
observations in the low-$z$ domain also for $q_o = 0.1$.

\end{document}